\documentclass[10pt]{iopart}
\usepackage{iopams,graphicx}

\newcommand{\beq}{\begin{equation}}
\newcommand{\eeq}{\end{equation}}
\newcommand{\bea}{\begin{eqnarray}}
\newcommand{\eea}{\end{eqnarray}}
\newcommand{\rem}[1]{ }

\newcommand{\ket}[1]{\left| #1\right>}
\newcommand{\V}{{\hat \mathcal{V}}}
\newcommand{\M}{{\hat \mathcal{M}}}

\newcommand{\prd}{Phys. Rev. D}
\newcommand{\apjl}{Astrophys. J. Lett.}
\newcommand{\physrep}{Phys. Reports}

\begin{document}
\rapid{``Evaporation'' of a flavor-mixed particle from a gravitational potential}
\author{Mikhail V. Medvedev}
\ead{mvm@ias.edu}
\address{
Institute for Advanced Study, Einstein Drive, Princeton, NJ 08540\\
Niels Bohr International Academy, Blegdamsvej 17, DK-2199 K\o benhavn \O, Denmark\\
Department of Physics and Astronomy, University of Kansas, Lawrence, KS 66045\\
Institute for Nuclear Fusion, RRC ``Kurchatov Institute'', Moscow 123182, Russia}

\begin{abstract}
We demonstrate that a stable particle with flavor mixing, confined in a gravitational potential can gradually and irreversibly escape -- or ``evaporate'' -- from it. This effect is due to mass eigenstate conversions which occur in  interactions (scattering) of mass states with other particles even when the energy exchange between them is vanishing. The evaporation and conversion are quantum effects not related to flavor oscillations, particle decay, quantum tunneling or other well-known processes. Apart from their profound academic interest, these effects should have tremendous implications for cosmology, e.g., (1) the cosmic neutrino background distortion is predicted and (2) the softening of central cusps in dark matter halos and smearing out or destruction of dwarf halos were suggested.
\end{abstract}

\section{Introduction} 

What happens to a particle trapped in a gravitational potential? The answer is obvious: it will remain gravitationally bound forever. The particle will never escape from the potential even if it interacts with other particles, provided these interactions preserve the total energy of the particle. Paradoxically, this answer turns out to be false for a flavor-mixed particle, such as a neutrino, for example. In this paper we show that a stable mixed particle, which bounces in a gravitational potential and scatters off other particles from time to time, gradually escapes --- or ``evaporates'' --- from it. Strictly speaking, the probability to detect the particle inside the potential decreases monotonically with time and, of course, the probability of its detection elsewhere increases. Such `evaporation' is a result of the conversion effect of mass eigenstates (in particular, of a heavier state into a lighter one) --- another process that has not been addressed. We underscore that particle evaporation and mass-state conversion processes have nothing to do with flavor oscillations, particle decay or quantum tunneling. 

Besides its profound academic and general physics interest, the evaporation effect shall have tremendous implications for cosmology. Almost one quarter of the energy and matter in the universe is dark matter, whose presence is revealed via its gravitational interaction only (see, e.g., Ref. \cite{KolbTurner}, for an overview), e.g., via gravitational strong and weak lensing, via stellar and gas kinematics, etc. The analysis of how the observed large-scale structure of the universe has formed supports the cold dark matter model with a cosmological constant ($\Lambda$CDM), in which non-relativistic (i.e., `cold') elementary particles gravitationally collapse to form dark matter filaments and halos. Gravity of the halos traps normal gas, which either collapses further to form stars and galaxies, or remains hot in galaxy clusters where it is detected with X-ray observations. Usually, dark matter dominates the mass of a self-gravitating system, e.g., the gas mass in clusters is only about 10\% of the total and the rest 90\% is dark matter. The origin of dark matter remains unknown, but the currently favored models, e.g., the axion dark matter and weakly interacting massive particles (WIMPs), admit them to be mixed particles: an axion can be mixed with a photon and WIPMs are a mixture of several supersymmetric particles (see, e.g., Ref. \cite{KG10,Bassan+10,Feng06,A-H09}, for recent reviews). How gravitational collapse of such flavor-mixed CDM occurs in the presence of evaporation and to what extent the formed dark halos and the entire large-scale structure of the universe are affected by this process is of paramount importance for modern precision cosmology. We suggest that dark matter evaporation may be able to address the outstanding problems of the CDM model: \cite{Kravtsov10,KdN+10} the overproduction of small-scale halos known as the `missing satellite problem' and the CDM prediction of diverging density profiles (cusps) in halo cores, not observed in dark-matter-dominated galaxies.

After nearly eighty years since its discovery and decades of observations and active search, the nature and properties of dark matter remain unknown. The possibility to constrain its mixing properties is very exciting; the study can readily be done with cosmological simulations.

Another intriguing cosmological implication deals with the cosmic neutrino background (CNB) --- the copious relic neutrino leftover from the big bang epoch \cite{Dolgov08}, whose present day temperature is $\lesssim2$~K. Although neutrino masses are unknown, an estimate \cite{Bilenky+02} of $m\sim(\Delta m^2)^{1/2}\sim0.1-0.01$~eV implies that cosmological neutrinos are nonrelativistic and can be trapped in CDM halos with escape velocities of the order of a few thousand km s$^{-1}$, so they can be affected by conversions and evaporation. Thus, if one would be able to detect CNB on Earth, one can notice that its flavor composition is noticeably skewed compared to the traditionally expected equipartition distribution.

\section{Description} 

The propagation (mass) and interaction (flavor) eigenstates of mixed particles 
are related by a unitary transformation, $\ket{f_i}=\sum_j U_{ij}\ket{m_j}$, where $\ket{f}$ and $\ket{m}$ denote the flavor and mass states, and $U$ is a unitary matrix. For simplicity, we consider a stable two-flavor particle, so $U$ is a $2\times2$ rotation matrix through an angle $\theta$ ($\theta$ being the mixing angle). The particle interacts in the flavor basis, so the interaction matrix is diagonal in this basis, $\tilde V=\textrm{diag}(V_\alpha,V_\beta)$, but it is non-diagonal in the mass basis, $V=U^\dagger \tilde V U$. The masses of the mass-states are $m_h$ and $m_l<m_h$, i.e., heavy and light. In general, mass states have different four-momenta and propagate along different geodesics.

The process of evaporation is easily explained with a specific example of an electron-muon neutrino system in one dimension.  Let's consider a thought experiment in which an electron neutrino is created inside a gravitational potential, $\phi(x)$. Energies and momenta of the mass states depend on the production process. By choosing them appropriately, we make the velocity of the light state to exceed the escape velocity of the potential, $v_{\rm esc}$, and the velocity of the heavy state to be smaller than $v_{\rm esc}$. Then, the trajectory of $\ket{m_h}$ is bound inside $\phi(x)$ and that of $\ket{m_l}$ is unbound and escapes to infinity (e.g., it is always so if $\ket{m_l}$ is relativistic or massless). In time, only the heavy state is left inside and if nothing else happens, this is the final state of the system.

Next, we let $\ket{m_h}$ to scatter off another particle (mixed or not) of mass $M$. Scattering of particles occurs through $V$, which is non-diagonal in the mass basis. The off-diagonal terms, $V_{lh}=V_{hl}=(V_e-V_\mu) \cos\theta\sin\theta$, couple different mass states, so that the amplitudes of  $\ket{m_h}$ and $\ket{m_l}$ can be non-zero upon scattering, i.e., the scattered wave is a mixture of mass states again. Hence, there are two channels $m_h\to m_h$ and $m_h\to m_l$, which are the elastic scattering of a mass-state and conversion of a larger-mass state into a lower-mass one. 

The total energy and momentum must be conserved in both processes. All is trivial for elastic scattering, so we consider the conversion process only. In the center of mass frame the momentum and energy conservations are $p_h+p_M=0=p_l'+p_M'$ and $(m_h^2+p_h^2)^{1/2}+(M^2+p_M^2)^{1/2}=(m_l^2+{p'_l}^2)^{1/2}+(M^2+{p'_M}^2)^{1/2}$, where `prime' means `after scattering' and $c=\hbar=1$ throughout, that is 
\beq
(m_h^2+p_h^2)^{1/2}+(M^2+p_h^2)^{1/2}=(m_l^2+{p'_l}^2)^{1/2}+(M^2+{p'_l}^2)^{1/2}.
\label{cons}
\eeq
Two special cases worth consideration: (a) $M\to\infty$ and (b) $m_h\approx m_l\approx M$. The first corresponds to the interaction with an external potential (an axion in a $B$-field) or with a heavy particle (a neutrino interacting with matter). The second is interesting for the self-interacting WIMP model ($\Delta m=m_h-m_l\ll m_l$ ensures that both supersymmetric particles are stable, see below). In both cases, we have $m_h^2+p_h^2\approx m_l^2+{p'_l}^2$, i.e., the kinetic energy of the secondary exceeds that of the primary. (Note that the reverse conversion process, $m_l\to m_h$ can, therefore, be kinematically suppressed.) In case (a), $\ket{m_l}$ can be relativistic with $p'_l\sim m_h$, if $m_h\gg m_l$, or non-relativistic with the velocity $v'_l\sim (2\Delta m/m_l)^{1/2}$, if $\Delta m\ll m_l$; in case (b) both mass states are nonrelativistic and $v'_l\sim (2\Delta m/m_l)^{1/2}$.

After the scattering, the two mass states propagate along the geodesics again, and if $v_l'>v_{\rm esc}$, then $\ket{m_l}$ leaves the potential whereas $\ket{m_l}$ remains trapped. Therefore, upon such a process the amplitude of the heavy state decreased irreversibly. The total probability is still unity, but the probability to detect a particle (an electron neutrino, for example) inside the potential has decreased and the probability of its detection somewhere outside has become larger. Of course, the overall energy is conserved: the light state climbs up the potential and looses energy (and a massless particle is redshifted). By repeating this cycle, one can further decrease the amplitude of the trapped state; colloquially speaking, the particle ``evaporates'' from the potential well.

We illustrate the effect numerically. We set $M\to\infty$ (hence, the velocity of the scatterer is negligible) and $\Delta m\ll m_l$ (hence, all particles are non-relativistic). The conservation of four-momentum, Eq. (\ref{cons}), in this case is
\beq
p_h^2/2m_h={p'_l}^2/2m_l-\Delta m.
\label{energy}
\eeq
To study the evolution of the non-relativistic mass states, we solve the two-component Schr\"odinger equation,
\beq
i\partial_t\ket{m(t,x)}=(H^{\rm free}+H^{\rm grav}+V)\ket{m(t,x)},
\eeq
where $\ket{m(t,x)}=\left(a_h(t,x),a_l(t,x)\right)^T$ is the state vector.
Here the free particle Hamiltonian, $H^{\rm free}={\rm diag}(-\partial^2_{xx}/2m_h,-\partial^2_{xx}/2m_l-\Delta m)$, satisfies condition (\ref{energy}). Gravity enters via $H^{\rm grav}={\rm diag}(m_h\phi(x),m_l\phi(x))$, where we chose a model attractive potential with exponential screening, $\phi(x)=\phi_0e^{-(x/x_g)^2}(1+(x/x_g)^2)^{-1}$, where $\phi_0<0$ determines its depth and $x_g$ sets its size ($x_g\sim0.5$ in computational units). For simplicity, we set $V_\mu=0$ and $V_e\not=0$. The scatterer is placed off-center, at $x_s\sim0.1$ and the scattering potential was chosen to be well-localized in space, $V_e=V_0e^{-((x-x_s)/x_v)^2}(1+((x-x_s)/x_v)^2)^{-1}$, where $V_0<0$ and $x_v\sim0.005$; the actual shape of $V(x)$ does not affect qualitative results so long as $x_v$ is sufficiently small. We solve the initial value problem with gaussian wave packets and different momenta of the mass states at $t=0$.

\begin{figure}
\flushright\includegraphics[width=0.9\columnwidth]{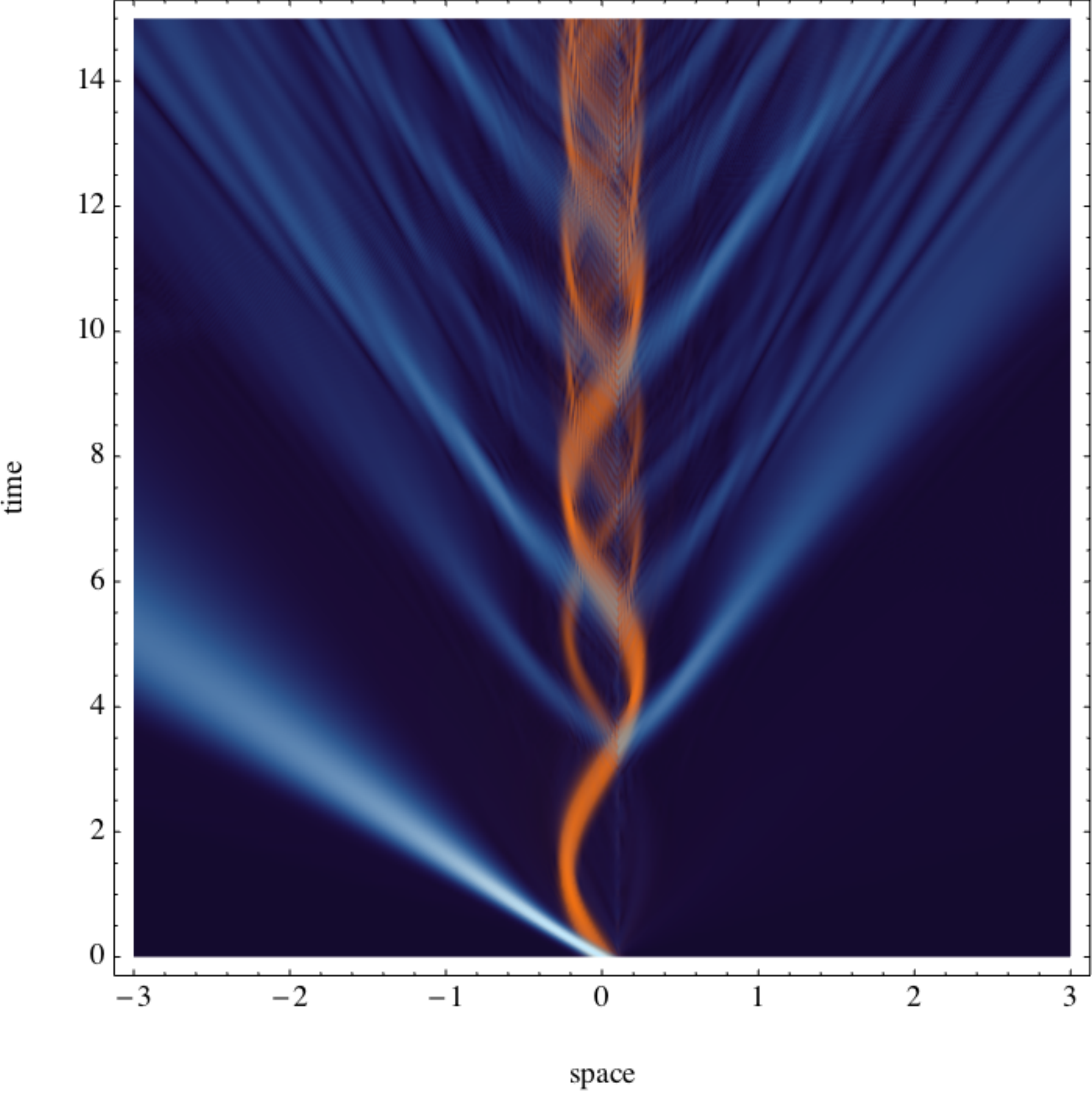}
\caption{Space-time diagram of a flavor-mixed particle in a gravitational potential obtained by solving a two-component Schr\"odinger equation. The probability densities of the light (blue) and heavy (orange) mass eigenstates are shown. The potential is localized between $x\sim-1$ and $+1$, scattering occurs at $x\sim0.1$. At $t=0$ a particle is created in a flavor state. The first scattering occurs at $t\sim3.5$. Each scattering produces forward and reflected wave packets; those corresponding to light mass eigenstates escape and heavy remain bound.}
\label{soln}
\end{figure}

The exact numerical solution of the Schr\"odinger equation of a mixed particle is shown in Fig. \ref{soln}. It represents the space-time diagram of the probability density (i.e., the amplitude squared) of the heavy (orange) and light (blue) mass states. (The amplitudes of flavor states can be obtained easily as a linear combination, $\ket{f_i}=\sum_j U_{ij}\ket{m_j}$.) Initially, the particle is in a flavor state. This is a coherent combination of two mass states, which propagate differently and, in time, the light state escapes. In contrast, the heavy state bounces off the potential and at $t\sim3.5$ scatters off the small-scale potential at $x=0.1$ for the first time. At this time, forward-scattered and reflected waves emerge, both are mixtures of two mass states. The light state escapes to infinity again but the heavy one keeps bouncing. Repetitive scatterings produce more outgoing light state wave packets at later times.

The purpose of this toy model solution was to demonstrate the existence of the evaporation and conversion processes. We believe it is a very beautiful result on its own. Applications to real physical systems may require modifications. For instance, for dark matter particles, the gravitational potential shall be computed self-consistently from the particle distribution.

\section{Extension and remarks} 

Gravitational field is efficient but not the only way to separate mass eigenstates. Other possibilities include scattering with gravitons (a rather weak, but not impossible process) or just ballistic propagation with different velocities. Thus, the above model of a gravitationally confined particle  can be reduced to a simpler ``particle-in-a-box" system, whose walls are membranes or semi-transparent mirrors, which keep $m_h$ inside but let $m_l$ to escape. We can formally define the membrane operator $\M$, which  separates out a light state (moves it out of the box) and does nothing to a heavy state, $\M(a_h^\downarrow\ket{m_h}+a_l^\downarrow\ket{m_l})=(a_h^\downarrow\ket{m_h}+a_l^\uparrow\ket{m_l})$. Similarly, scattering is described by $\V$-operator, $\V(a_l^\downarrow\ket{m_l})=({a'_h}^\downarrow\ket{m_h}+{a'_l}^\downarrow\ket{m_l})$, where up and down arrows label the amplitudes localized outside and inside the box, respectively. Then the evaporation process reads as ``separation-scattering-separation'', or $\M\V\M$, and 
\beq
\M\V\M(a_h^\downarrow\ket{m_h}+a_l^\downarrow\ket{m_l})=
{a'_h}^\downarrow\ket{m_h}+(a_l^\uparrow+{a'_l}^\uparrow)\ket{m_l}.
\eeq
The process shown in Fig. \ref{soln} can be represented as an operator $(\M\V)^n\M=(\M\V)\dots(\M\V)\M$, where $n$ is the number of scatterings. This formalism allows an easy calculation of the amplitudes, ${a'}_{h,l}^{\dots}\equiv a_{h,l}^{(n)}$, as a function of $n$ or time. In particular, $|a_{h}^{(\infty)}|^2=0$ and $|a_{l}^{(\infty)}|^2=1$, that is, a non-decaying  particle can be entirely converted into a light mass state and become unconfined. 

Note that scatterings in the box shall not happen too frequently, because the mass states have to separate; otherwise they interact coherently as a well-defined flavor state.
For completeness, we qualitatively consider two more cases. In case (A), let's take a box that is completely sealed (no membranes), so both mass states are kept inside. Then an equilibrium shall exist and two regimes are possible. First, if the wave packets are wide enough (it depends on production/detection, reflection and scattering), then they overlap and scatter coherently, hence transition probabilities shall be oscillatory functions of space and/or time. Therefore, equilibrium can depend on masses and four-momenta of mass states, the box size, $L_\mathrm{box}$, and even parameters of the scattering potential. Second, when the overlap of wave packets vanishes, scattering of mass states occurs independently, so an equilibrium is set by the detailed balance, $\ket{m_h}\rightleftarrows\ket{m_l}$. Equality of the rates of the forward and reverse processes determines $a_i$ --- the mass and flavor composition of the system. In case (B), we put our original ``particle-in-a-box'' system into a larger sealed box of size $L_\mathrm{box,2}>L_\mathrm{box}$ and consider the incoherent regime only (note that light states can come back into the smaller box, but heavy states cannot go out of it). The mass conversion is now skewed toward  $\ket{m_h}\to\ket{m_l}$ process, because the rate of the opposite process is reduced by a factor of $(L_\mathrm{box}/L_\mathrm{box,2})^{1/2}$ (since the wave functions are normalized by box volumes). Hence, the composition is also skewed towards the light state. In the limit of $L_\mathrm{box,2}\to\infty$, one recovers $|a_h^{(\infty)}|^2=0$.

\section{Implications} 

Possible important implications mentioned in Introduction are estimated here.  A more detailed analysis is beyond the scope of the paper.

First, cosmological neutrinos of the CNB. Assuming a nominal non-relativistic neutrino cross-section \cite{Shvartsman+82} $\sigma_\nu\sim10^{-60}\textrm{ cm}^2$, the mean density of nucleons in halos $n\sim10^{-3}\textrm{ cm}^{-3}$, and the random velocity of $v\sim10^{-3}$, we estimate the number of scatterings to be $\sim n\sigma_\nu v t_H\sim10^{-47}$ per neutrino for the life-time of the universe (the Hubble time), $t_H=4\times10^{17}$~s. The effect is very tiny and unobservable (unless nonrelativistic neutrinos (self-)interact more strongly than we currently think), though it may have some effect for the ultimate fate of the Universe, at $t_H\to\infty$. However, $\sigma_\nu$ is tremendously enhanced by $\sim N^2 Z^2$ for coherent scattering, where $Z$ is the charge of the atomic nucleus and $N\sim n\lambda^3$ is the number of nuclei within volume $\sim \lambda^3$ and $\lambda=1/mv$ is the neutrino wavelength \cite{Shvartsman+82}. Assuming $\lambda\sim1$~cm, for Earth ($n\sim10^{23}\textrm{ cm}^{-3}$, nominal $Z\sim25$ and distance $L\sim10^4$~km) the conversion ``optical depth" is $\tau_c\sim10^{-4}$, that is CNB passing through Earth is slightly affected by conversions.

Some models of dark matter involve a hypothetic sterile neutrino, which is presumably more massive than the known ones. The results of the evaporation model can be readily generalized to the sterile neutrino scenario.

In the axion CDM model, axions can scatter off cosmic magnetic fields. Using the experimental upper limit on axion-photon coupling $g_{a\gamma}\sim10^{-11}\textrm{ GeV}^{-1}$, we estimate the interaction/conversion probability \cite{KG10,Bassan+10} to be $\sim 0.1(g_{a\gamma}/10^{-11}\textrm{ GeV}^{-1})(B/1~\mu{\rm G})(L/10~{\rm kpc})$. For galactic halos with a typical galactic field strength of, say, $B\sim1-3~\mu{\rm G}$ and the halo sizes $L\sim30-100$~kpc, the probability is of order unity per passage. In galaxy clusters with $B\sim0.1~\mu$G and sizes of a couple of Mpc, the rate can be similar. Note that `one-per-passage' is an efficient regime of conversions, so the cosmological effect can be noticeable. If occurs, it can lead to substantial evaporation of axion CDM halos on the Hubble time-scale. If the effect is not observed, it can further constrain the coupling constant, $g_{a\gamma}$, or rule out axion-CDM.

The WIMP CDM model deals with the lightest supersymmetric particle (neutralino), higher-mass ones are assumed to decay quickly. However, for high mass degeneracy, which we consider now, all decay channels can be kinematically forbidden, so more than one particle can be stable. Taking rather crudely $\Delta m\sim\textrm{ MeV}$ and $m\sim\textrm{ TeV}$, we estimate the velocity of a produced light state to be $v'_l\sim(\Delta m/m)^{1/2}\sim300\textrm{ km s}^{-1}$, which is a cosmologically interesting number. It is smaller or comparable to the escape velocity from halos of large galaxies and galaxy clusters, but it is larger then $v_{\bf esc}$ for halos of dwarf galaxies, which can be as low as a few tens km~s$^{-1}$. A halo dichotomy is then expected: the number of small halos must be smaller than what CDM simulations predict whereas the population of large halos is mainly unaffected. Such an effect can be an alternative explanation to the `missing satellite problem' and, perhaps, it can also affect halo cores \cite{Kravtsov10,KdN+10}. Indeed, mass-conversion is more efficient at higher densities (more scatterings). Therefore, higher evaporation rate from high-density environments can destroy or smear out small high-concentration halos and smooth out the mass distribution in the cores. 

For the effect of conversions to be significant, the rate shall be at least a few reactions per Hubble time. Dark matter self-interaction can accommodate the needed large cross section, though simple models seem to be disfavored \cite{Kravtsov10,KdN+10}. The models with the velocity-dependent cross-section and the Sommerfeld enhancement \cite{A-H09}, which is the most profound at low velocities (that is, in dwarfs, again), shall be tested before this scenario is ruled out entirely. Meanwhile, a nice feature of our model is that elastic cross-sections for self-interacting CDM and for CDM scattering off normal matter, $M$, can be made as small as desired (to comply with observational constraints), while keeping the conversion cross-section large. For instance, for scattering $h+M\to h+M$ and $l+M\to l+M$, for $\theta\sim\pi/4$ and $V_\alpha\sim-V_\beta$, one readily has $V_{hh}\sim V_{ll}\sim0,\ V_{lh}=V_{hl}\sim V_\alpha$. The disadvantage of this scenario is that it required strong degeneracy, which needs an explanation. However, a similar but {\it ad hoc} model with GeV-scale WIMPs with a keV-scale degeneracy seems to explain some direct detection experimental data and spacecraft flyby anomalies \cite{Essig+10,Adler09}.

To conclude, we have demonstrated that elementary particles with flavor mixing, like neutrino, kaons, etc., that are initially trapped in the external gravitational field can gradually escape from it in the process involving conversions of mass states. In retrospect, the process has a simple interpretation. In ordinary non-relativistic quantum physics one typically chooses the zero of energy such that bound states have negative energy. In relativistic quantum physics it is more correct to say that bound states correspond to quantum states with an energy that is less than the rest energy of the particle in question (assuming the potential asymptotes to zero at large distances).  In a flavour-mixed system, if the energy of the ``bound state" is less than the rest energy of the initial mass eigenstate the system is prepared in, but greater than the rest-energy of any one of the other mass eigenstates the particle can mix into, then there will be some probability of escape. For the escape process to proceed one needs a flavor-mixed system to experience interactions in the basis that is not aligned with the propagation (mass) basis. In this paper we also considered a number of possible implications of the conversion effect for the cosmic neutrino background and dark matter. We also argue that if dark matter in the universe is a flavor-mixed particle, the suggested evaporation process can resolve several outstanding problems of the modern cosmology related to the large-scale structure formation of the Universe.

\ack
The author thanks Nima Arkani-Hamed and Peter Goldreich for thoughtful discussions, Scott Tremaine and Masataka Fukugita for interesting suggestions, Freeman Dyson and Stephen Adler for encouragement. 
The author acknowledges support from The Ambrose Monell Foundation (IAS) and The Ib Henriksen Foundation (NBIA), as well as grants AST-0708213 (NSF), NNX-08AL39G (NASA) and DE-FG02-07ER54940 (DOE).


\section*{References}
\begin{thebibliography}{10}

\bibitem{KolbTurner}
{Kolb} E W and {Turner} M S 1990
\newblock {\em The early universe}
\newblock (Reading: Addison-Wesley)

\bibitem{KG10}
{Kim} J E and {Carosi} G 2010
\newblock {\em Reviews of Modern Physics} {\bf 82} 557

\bibitem{Bassan+10}
{Bassan} N, {Mirizzi} A and {Roncadelli} M 2010
\newblock {\em JCAP} {\bf 05} 010

\bibitem{Feng06}
{Feng} J L 2006
\newblock {\em Journal of Physics G Nuclear Physics} {\bf 32} R1

\bibitem{A-H09}
{Arkani-Hamed} N, {Finkbeiner} D P, {Slatyer} T R and {Weiner} N 2009
\newblock {\em \prd} {\bf 79} 015014

\bibitem{Kravtsov10}
{Kravtsov} A 2010
\newblock {\em Advances in Astronomy} {\bf 2010} 1

\bibitem{KdN+10}
{Kuzio de Naray} R, {Martinez} G D, {Bullock} J S and {Kaplinghat} M 2010
\newblock {\em \apjl} {\bf 710} L161

\bibitem{Dolgov08}
{Dolgov} A D 2008
\newblock {\em Physics of Atomic Nuclei} {\bf 71} 2152

\bibitem{Bilenky+02}
{Bilenky} S M, {Giunti} C, {Grifols} J A and {Mass{\'o}} E 2003
\newblock {\em \physrep} {\bf 379} 69

\bibitem{Shvartsman+82}
{Shvartsman} V F, {Braginski{\v \i}} V B, {Gershte{\v \i}n} S S, {Zel'dovich} Y B and {Khlopov} M Y 1982
\newblock {\em Soviet Journal of Experimental and Theoretical Physics Letters} {\bf 36} 277

\bibitem{Essig+10}
{Essig} R, {Kaplan} J, {Schuster} P and {Toro} N 2010
\newblock {\em ArXiv e-print}: 1004.0691

\bibitem{Adler09}
{Adler} S L 2009
\newblock {\em \prd}, {\bf 79} 023505

\end{thebibliography}
\end{document}